\documentclass[aps,pra, superscriptaddress, nofootinbib,longbibliography]{revtex4-2}

\usepackage[utf8]{inputenc}
\usepackage{amsmath, amssymb}
\usepackage{graphicx}
\usepackage{mathrsfs} 
\usepackage[colorlinks=true, allcolors=blue]{hyperref}
\usepackage{braket} 

\begin{document}

\title{Neutral-atom quantum computation using multi-qubit geometric gates via adiabatic passage}

\author{Sinchan Snigdha Rej and Bimalendu Deb }
\email{msbd@iacs.res.in} 
 \affiliation{School of Physical Sciences, Indian Association for the Cultivation of Science, Jadavpur, Kolkata 700032, India.}

\date{\today}

\begin{abstract}
   Adiabatic geometric phase gates offer enhanced robustness against fluctuations compared to conventional Rydberg blockade-based phase gates that rely on dynamical phase accumulation.  We theoretically demonstrate two- and multi-qubit phase gates in a neutral atom architecture, relying on a double stimulated Raman adiabatic passage (double-STIRAP) pulse sequence that imprints a controllable geometric phase on the qubit systems. The system is designed in such a way that every atom is individually addressable, and moreover, no extra laser is required to be applied on the target atom while scaling up the system from two- to multi-qubit quantum gates. The gate fidelity has been numerically analyzed by changing the gate operation time, and we find that $98\%$ to $99\%$ fidelity can be achieved for gate time $\simeq0.6\ \mu$s.  We perform a systematic error analysis, which reveals that our proposed gates can exhibit strong resilience against fluctuations in Rabi frequencies, finite blockade strength, and atomic position variations. These results establish our approach as a physically feasible and scalable pathway toward fault-tolerant quantum computation with neutral atoms. We simulate Grover’s search algorithm for two-, three-, and four-qubit systems with high success probability and thereby demonstrate the utility and scalability of our proposed gates for quantum computation.

\end{abstract}

\maketitle

\section{Introduction} Rydberg atom systems have emerged as a leading quantum computing platform over the past two decades, due to their unique features, such as long-range interactions and scalability relative to other quantum architectures \cite{PhysRevLett.85.2208, Saffman2010Quantum}. A defining characteristic of this system is the Rydberg blockade, when one atom is excited to a Rydberg state, excitations of nearby atoms within a certain radius are energetically suppressed. Exploiting this mechanism, researchers have developed a variety of gate protocols \cite{PhysRevLett.104.010503, PhysRevA.92.022336, PhysRevA.94.032306, PhysRevApplied.7.064017, PhysRevA.104.012615, PhysRevA.110.062618}, including global laser pulses\cite{PhysRevLett.123.170503, Evered2023HFPE}, optimally shaped analytic pulses\cite{maller2017high}, and protocols based on adiabatic dark-state evolution \cite{Mueller2009Mesoscopic, PhysRevLett.129.200501, RejDeb2025_ToffoliCnNOT}. Beyond blockade, the Rydberg antiblockade regime allows simultaneous excitation of multiple atoms within the blockade radius under specific detuning and pulse conditions. This effect is used to devise alternative entangling gate schemes \cite{PhysRevA.95.022319, Wu2021AntiblockadeSWAP, Wu2021ResilientRydberg,  Li2023HighToleranceAntiblockade}. Quantum logic gates based on Rydberg blockade invariably suffer from blockade‐induced errors that scale as $(\Omega/V)^2$, where $\Omega$ and $V$ are the single‐atom Rabi frequency and the Rydberg–Rydberg interaction strength, respectively \cite{Saffman2010Quantum}. While one may suppress these errors by increasing $V$, doing this leads to unwanted mechanical forces on the atoms and thus introduces additional decoherence pathways. Alternative strategies for minimizing blockade errors include the use of optimized, generalized Rabi pulses that cancel leading‐order infidelities \cite{Shi2017RydbergFreeBlockade, Su2023RabiBlockadeAllGeometric}, as well as protocols that exploit dark‐state dynamics involving Rydberg levels to decouple the system from loss channels \cite{Mueller2009Mesoscopic}. Beyond blockade imperfections, fluctuations in the driving laser intensity give rise to Rabi errors that become especially pronounced at high $\Omega$ and likewise limit gate fidelity \cite{Su2023RabiBlockadeAllGeometric, JinJing2024_GeometricDarkPaths}. One can face another issue of positional fluctuation of atoms while scaling up the system. An ideal neutral‐atom quantum processor must therefore address blockade, Rabi, and positional fluctuation errors in concert.  

The geometric phase \cite{Ekert2000_GeometricQuantumComputation} depends solely on the global features of the evolution path traced during a cyclic process, rather than on local dynamical details. Leibfried \emph{et al.} first experimentally demonstrated a high-fidelity geometric two-ion-qubit phase gate based on coherent displacements induced by an optical dipole force \cite{Leibfried2003}. Quantum logic gates constructed using such geometric phases (holonomies) thus possess intrinsic robustness against local control noise and parameter fluctuations \cite{Wu2010_ControlledGeometricPhaseGate, JinJing2024_GeometricDarkPaths, Chen2024_SinglePulseNoncyclic, Zhang2022_GroundStateBlockadeGate, Jin2024_NewTypeGeometricGates}. Among various implementations, adiabatic protocols are particularly attractive for large-scale quantum computation due to their inherent resilience to control imperfections and parameter drifts \cite{ Rao2014_RobustRydbergAdiabatic, Beterov2016_StarkTunedForster, Saffman2020_SymmetricRydbergCZ, Wu2018_RydbergAdiabaticPhaseControl, Liao2019_GeometricRydbergSTA, Zhou2020_STA_CZ_Rydberg}. A key advantage of these adiabatic Rydberg-gate schemes is the suppression of transient population in the Rydberg excited states. By maintaining adiabatic evolution, the system predominantly occupies the lower-energy dressed states, thereby minimizing spontaneous emission and other decay channels associated with Rydberg excitation. Consequently, the protocol offers enhanced protection against decoherence and yields improved gate fidelity. Entanglement generation via  STIRAP and the adiabatic evolution of a two-atom dark state within the Rydberg blockade regime has been analyzed by Møller \emph{et al.} \cite{Moller2008_RydbergAdiabaticPassage}. An interesting scheme using double-STIRAP (d-STIRAP) has been proposed earlier in the paper \cite{Moller2007}, demonstrating a geometric phase gate in a tripod-level system through the adiabatic evolution of a dark state around a closed loop in parameter space. Inspired by this approach, we propose a protocol for implementing multi-qubit controlled geometric phase (C$^n$Z) gates based on the d-STIRAP mechanism.

In this work, we propose an alternative pathway to realize a two-qubit geometric phase gate based on dark-state dynamics, employing a pair of d-STIRAP pulse sequences. The protocol begins with a soft $\pi$-pulse applied to the control atom, coupling one of its hyperfine ground states (qubit states) to a Rydberg level. Subsequently, the target atom undergoes a pair of d-STIRAP sequences with parametrized Rabi frequencies. In the presence of a longitudinal magnetic field, we choose three Zeeman sublevels in the target atom, two of which encode the qubit states while the third serves as an auxiliary level required for completing the d-STIRAP process. One qubit state is placed in two-photon resonance with the Rydberg state through an intermediate level, enabling a controllable d-STIRAP process with the auxiliary state that imparts a conditional geometric phase on the target qubit. The other qubit state undergoes a complete d-STIRAP cycle with the auxiliary state via a distinct intermediate level that is not coupled to any Rydberg excitation, thereby acquiring a geometric phase independent of the control state. A second soft $\pi$-pulse is then applied to the control atom to close the sequence and complete the gate operation. This scheme is further generalized to construct three- and four-qubit geometric phase gates in one- and two-dimensional atomic configurations, respectively. We design our model in such a way that we have no need to apply any extra laser on the target atom while scaling up from two- to multi-qubit gates.

We investigate the dependence of the amplitudes of the computational basis states on the system parameters. We include spontaneous decay of the Rydberg states, and analyze the fidelity of the two-, three-, and four-qubit phase gates as a function of gate operation time. We find that for gate time $\simeq0.6\ \mu$s, more than $98\%$ gate fidelity can be achieved.  We study the robustness of our proposed gates against systematic errors such as fluctuations in Rabi frequencies, finite blockade strength, and positional fluctuations. Our results show these multi-qubit gates are significantly more resilient compared to the conventional EIT-based CNOT gate \cite{Mueller2009Mesoscopic}. The numerical simulations are performed by solving the Lindblad master equation using QuTiP \cite{qutip2012}. Finally, to benchmark the computational utility of the scheme, we simulate Grover’s search algorithm with two, three, and four qubits, using our proposed multi-qubit phase gates. The corresponding circuits are executed in Qiskit \cite{Qiskit}, yielding high success probabilities and confirming the practical viability of our approach. 

The paper is organized as follows. In Sec.\ref{2Q} we introduce the protocol for implementing a two-qubit geometric phase gate. In Sec.\ref{MQ} this scheme is generalized to construct three- and four-qubit geometric phase gates. In Sec.\ref{results} we present numerical simulations of the computational basis amplitudes under experimentally relevant parameters and in Sec.\ref{fid} analyze the gate fidelity as a function of total gate operation time in the presence of spontaneous decay. Sec.\ref{errors} is devoted to an error analysis, where we examine the robustness of the proposed gates against systematic imperfections, including Rabi frequency fluctuations, finite blockade strength, and positional fluctuations. Finally, in Sec.\ref{grover} we demonstrate the application of the proposed gates to Grover’s search algorithm, thereby establishing their utility for quantum computing.

\section{construction of geometric phase gates} 
\subsection{Two-qubit}
\label{2Q}
 \begin{figure}
     \centering
    \includegraphics [width=12cm]{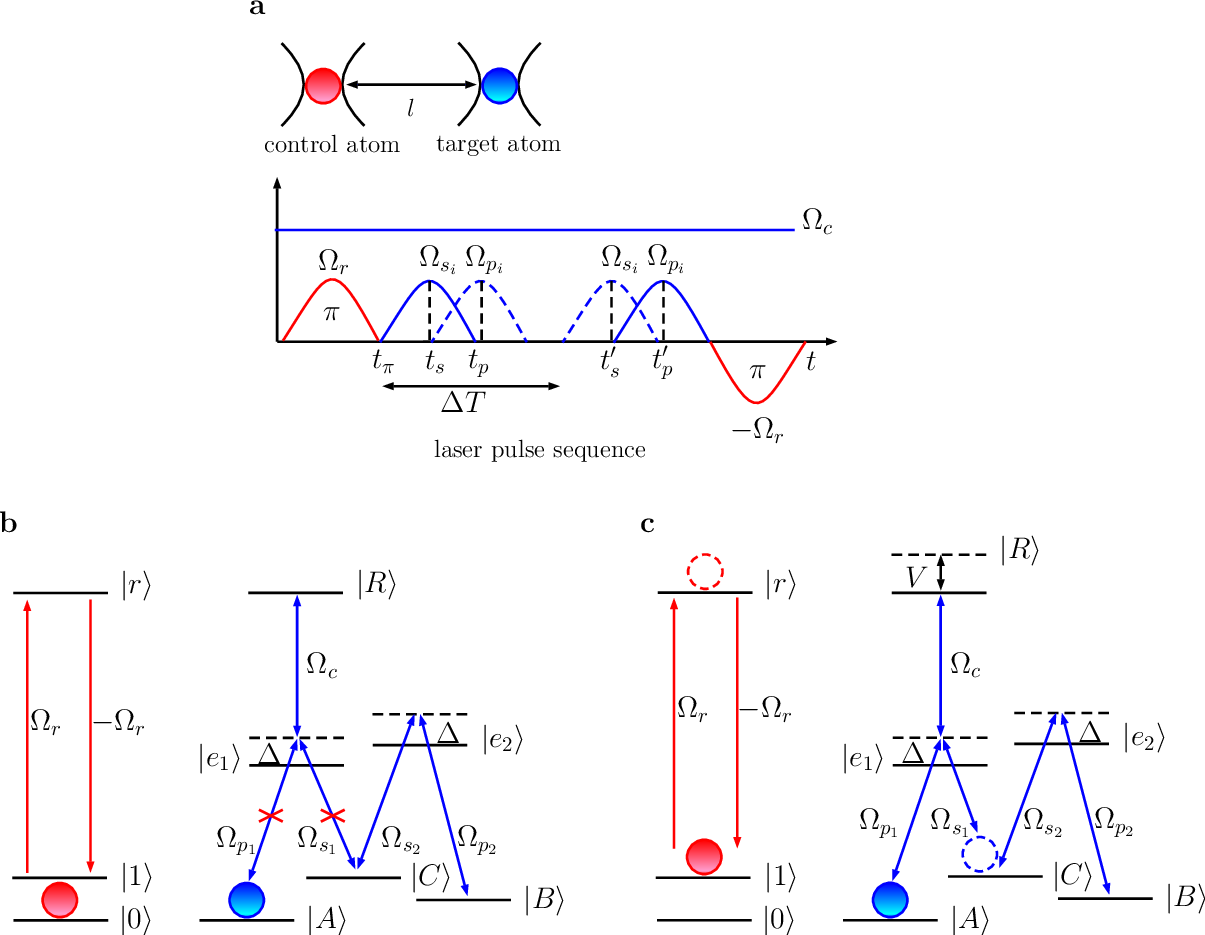}
     \caption{(a) The control and target atoms are arranged in two different optical traps with a fixed distance $l$ between the trap centers. The laser pulse sequence to complete the gate protocol is shown. (b) When the control atom is at $\ket 0$, the $\ket r$ state is not populated. As a result, there is no shift in the state $\ket R$, the dark state condition forbids any population transfer from the state $\ket{A}$. (c) If the control atom is in $\ket 1$, after the first $\pi$-pulse is applied, $\ket r$ gets populated, resulting in a shift $V$ in $\ket R$. This breaks the dark state condition, initiating population transfer from $\ket A$. After the d-STIRAP process is completed, the state gathers a geometric phase $\Gamma$ as described in the main text. Applying the second $\pi$-pulse on the control atom, we complete the gate protocol. }
     \label{fig1}
 \end{figure}
Let us consider that the control and target atoms are trapped in two different optical traps with a distance $l$ between the trap-centers, as shown in Fig.\ref{fig1}a. Two hyperfine ground states $\ket{0}$ and $\ket{1}$ constitute the qubit states with $\ket r $ being a Rydberg state of the control atom. For the target atom, we consider three Zeeman sublevels $\ket A$, $\ket B$, and $\ket C$, which are shifted due to an applied magnetic field. Here $\ket A$ and $\ket{B}$ are qubit states and $\Ket R$ is the Rydberg state of the target atom. Moreover, we make use of two excited intermediate states $\ket{e_1},\ket{e_2}$ in the target atom, as shown in Fig.\ref{fig1}, to complete the d-STIRAP process. The state $\ket 1$ is resonantly coupled to $\ket r$ with a soft $\pi$ pulse of amplitude $\Omega_r$ while $\ket0$ remains totally decoupled. We apply two pairs of Raman lasers to the target atom that couple the intermediate state $\ket{e_1}(\ket{e_2})$ to $\ket C$ and $\ket A(\ket{B})$ labelled by $S_1(S_2)$ (Stokes) and $P_1(P_2)$ (pump) lasers, respectively with a single-photon blue detuning $\Delta$. $S_i$ and $P_i$ have time-varying Rabi frequencies, $\Omega_{s_i}'(t)=\Omega_s(t)e^{-i\phi(t)}$ and $\Omega_{p_i}(t)$, both of which have an amplitude $\Omega_0$. The detunig $\Delta\gg\gamma_e$ to minimize the spontaneous decay from $\ket {e_i}$, when $\ket {e_i}$ has a lifetime $\gamma_e^{-1}$. A separate laser $\mathcal{L}$, characterized by a Rabi frequency $\Omega_c$ couples $\ket R$ and $\ket {e_1}$.

The pulse sequence for the gate protocol is shown in the Fig.\ref{fig1}a. First, we apply a soft $\pi$-pulse with a Rabi frequency of $\Omega_r$ on the control qubit defined by,
\begin{equation}
     \Omega_r(t)=\Omega_r/2(1-\cos(2\pi t)/\mathscr{T}), 
 \end{equation}
where $\mathscr{T}$ is the time period of the pulse. Then it is followed by a simultaneous pair of d-STIRAP pulse sequences on the target atom. 
\begin{equation}
\begin{aligned}
\Omega_{p_i}(t) &= \Omega_0/2  [ G(t,t_{p})  +  G(t,t_{p'})  ], \\
\Omega_{s_i}(t) &= \Omega_0/2  [  G(t,t_{s}) +  G(t,t_{s'})  ],
\end{aligned}
\end{equation}
where the Gaussian envelope is
\begin{equation}
G(t,t_0) = \exp \left[- {(t-t_0)^2}/{2\sigma^2} \right],
\end{equation}
$t_{p},t_{p'}$ and $t_{s},t_{s'}$ are described in Fig.\ref{fig1}a, $\sigma$ is the standard deviation of the Gausian pulse. For simplicity, we take $\Omega_0=\Omega_r$. Finally, another soft $\pi$-pulse is applied on the control qubit but with a Rabi frequency of $-\Omega_r$ to complete the gate operation.  Here we discuss the gate operation step by step.
\paragraph{Step-I}: The initial state is $\ket{0A}$. This step is illustrated in Fig.\ref{fig1}b. The first $\pi$-pulse on the control atom does not change its state, i.e. $\ket r$ remains unpopulated, and so there is no shift in the state $\ket R$. Now, during the d-STIRAP pulse sequences, the target atom will follow the Hamiltonian ($\hbar=1$),
\begin{equation}
    \mathcal{H}^1=\Omega_{s_1}'(t)/2 \ket{C}\bra{e_1}+\Omega_{p_1}(t)/2\ket{A}\bra{e_1}-\Delta\ket{e_1}\bra{e_1}+\Omega_c/2\ket{e_1}\bra{R}+\text{H.c.}
\end{equation}
for both halves of the process.
The above Hamiltonian has the following two dark states 
\begin{equation}
\begin{aligned}
     \ket{D_1}&={N(t)}^{-1/2}[\Omega_{s_1}'(t)\ket{A}-\Omega_{p_1}(t)\ket{C}]\\
   \ket{D_2}&=[N(t)+x(t)^2]^{-1/2}[\Omega_{p_1}(t)\ket{A}+\Omega_{s_1}'^*(t)\ket{C}-x(t)\ket{R}],
\end{aligned}
\end{equation} where $N(t)=(\Omega_{p_1}(t)^2+ \Omega_{s_1}(t)^2) $ and $x(t)= [(\Omega_{p_1}(t)^2+\Omega_{s_1}(t)^2)/\Omega_c]$. Satisfying the condition $\Omega_c/\Omega_0>2$ the target qubit starts evolving following the dark state $\ket{D}={N(t)^{-1/2}}(\Omega_{s_1}'\ket{D_1}+\Omega_{p_1}\ket{D_2)}$, which results in no change in the target state. After the completion of the second $\pi$-pulse on the control atom, we get the transformation $\ket{0A}\longrightarrow\ket{0A}$.

 \paragraph{Step-II}: The second step is illustrated in fig. \ref{fig1}c. We start from the initial state $\ket{1A}$. After applying the first $\pi$-pulse on the control atom, the state evolves to $-i\ket{rA}$, resulting in a shift in $\ket R$ by an amount of $V$ due to the Rydberg-Rydberg interaction. The breakdown of the dark state condition allows population transfer from $\ket A$ to $\ket C$ and vice versa, governed by the Hamiltonian
\begin{equation}
   \mathcal{H}^2=\mathcal{H}^1+V\ket{R}\bra{R} .
\end{equation}
for both halves of the d-STIRAP process.
After this process is over, the state $\ket A$ will come back to the same state but with a geometric phase, as described in \cite{Moller2007}
\begin{equation}
    \Gamma=\phi(t_\pi)-\phi(t_\pi+\Delta T),
\end{equation}
 where $t_\pi$ and $\Delta T$ are shown in the Fig.\ref{fig1}a. After the operation of the second $\pi$-pulse on the control qubit, we achieve our desired transformation, $\ket{1A}\longrightarrow e^{i \Gamma}\ket{1A}$.

\begin{figure}
     \centering
    \includegraphics [width=12cm]{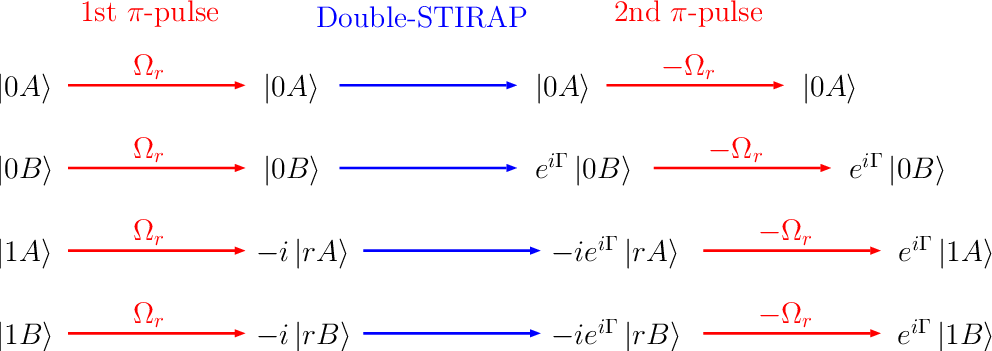}
     \caption{Schematic of the evolutions of all four initial states.}
     \label{fig2}
 \end{figure}

For the rest of the two initial states, i.e. $\ket{0B}$ and $\ket{1B}$, the target qubit will accumulate a geometric phase $\Gamma$ following the d-STIRAP process irrespective of the initial state of the control atom, as the intermediate state $\ket{e_2}$ is not coupled to the Rydberg state $\ket R$. For $\Gamma=\pi$, the $CZ$ gate can be realized. The full process of the gate operation is schematically shown in the Fig.\ref{fig2}.

  \subsection{Multi-qubit} 
\label{MQ}
\begin{figure}
     \centering
    \includegraphics [width=8cm]{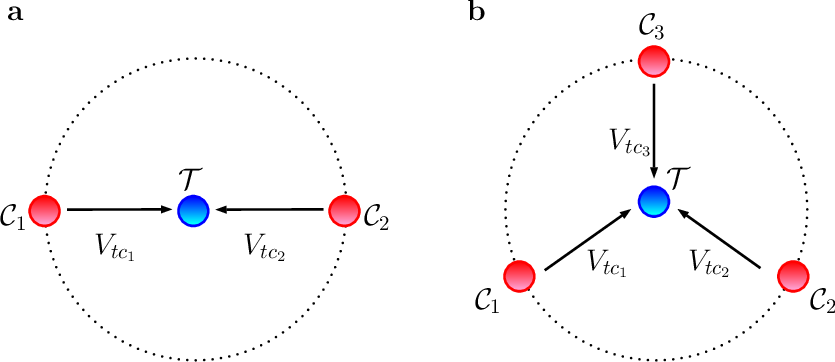}
     \caption{(a),(b) Atomic configuration for three- and four-qubit controlled phase gates, respectively. }
     \label{fig3}
 \end{figure}

To construct multi-qubit geometric phase gates, we can arrange the control and target atoms as shown in Fig.\ref{fig3}. For the three-qubit case, first, we will apply simultaneous soft $\pi$ pulses on both control atoms. When both control atoms are in $\ket{0}$, $\ket r$ will not be populated, resulting in no shift in $\ket{R}$ of the target atom. As the dark state condition holds, no evolution in the target atom will occur if it is in the state $\ket{A}$, $\ket{00A}\longrightarrow\ket{00A}$. If at least one of the control atoms is in $\ket{1}$, this will result in population transfer to $\ket{r}$ after the first $\pi$ pulse is applied, and the state $\ket{R}$ will be shifted due to Rydberg-Rydberg interaction. This shift breaks the dark state condition to start an evolution in the target atom. After the d-STIRAP process, the target state will accumulate a phase $\Gamma$. If the target atom is at $\ket{B}$, after the d-STIRAP process, the geometric phase $\Gamma$ will always appear irrespective of the initial state of the control atom. For $\Gamma=\pi$, the full gate operation can be represented by $
\mathrm{diag}(1,-1,-1,-1,-1,-1,-1,-1)
$. Similarly, for the four-qubit gate, when all control atoms are at $\ket{0}$ and the target atom is at $\ket{A}$, the population transfer will be blocked. For the rest of the initial states, a geometric phase will be accumulated, resulting in a four-qubit geometric phase gate.
 
 \section{Results and Discussions}
\label{results}

\begin{figure}
     \centering
     \includegraphics [width=8.5cm]
     {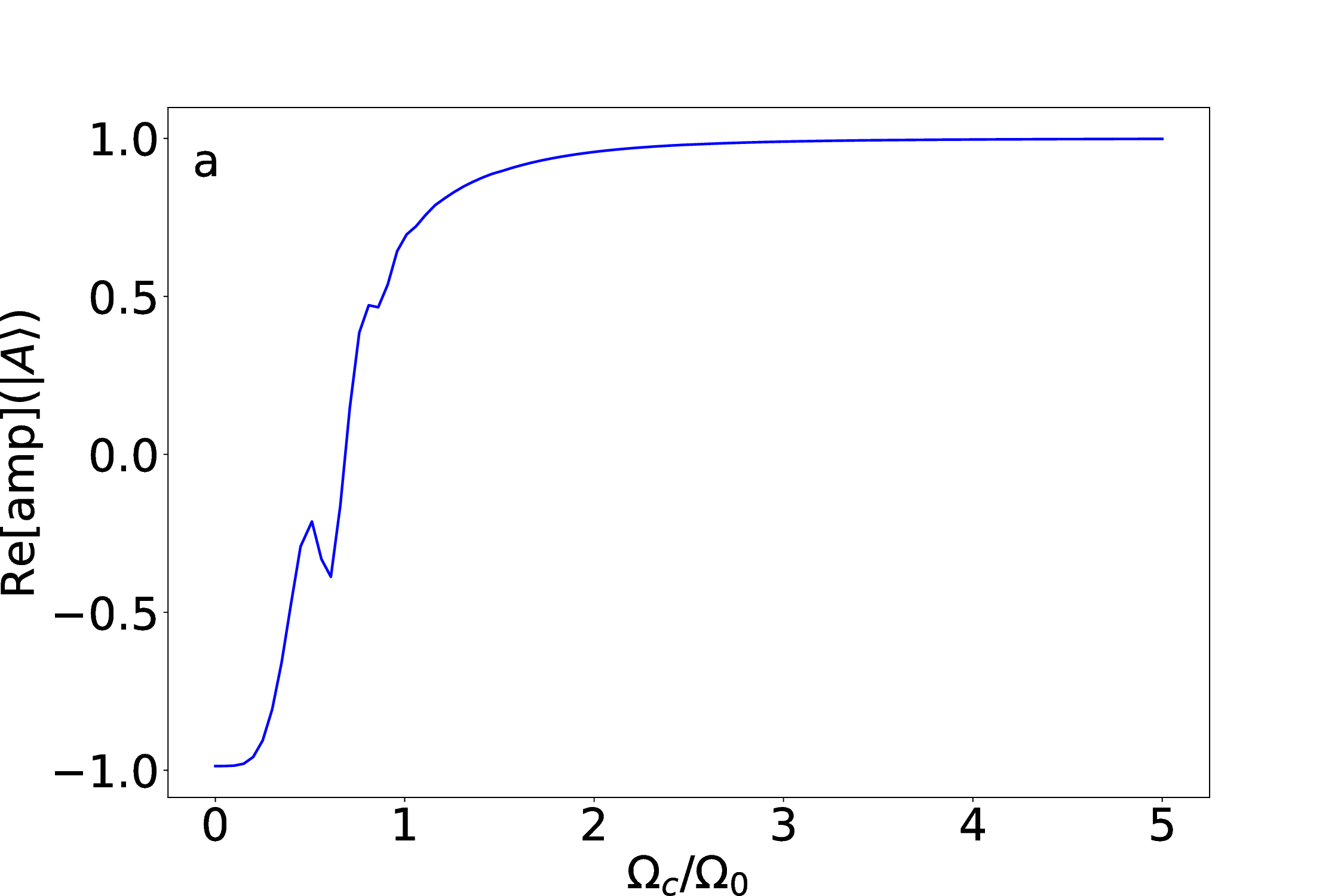}
     \includegraphics [width=8.5cm]{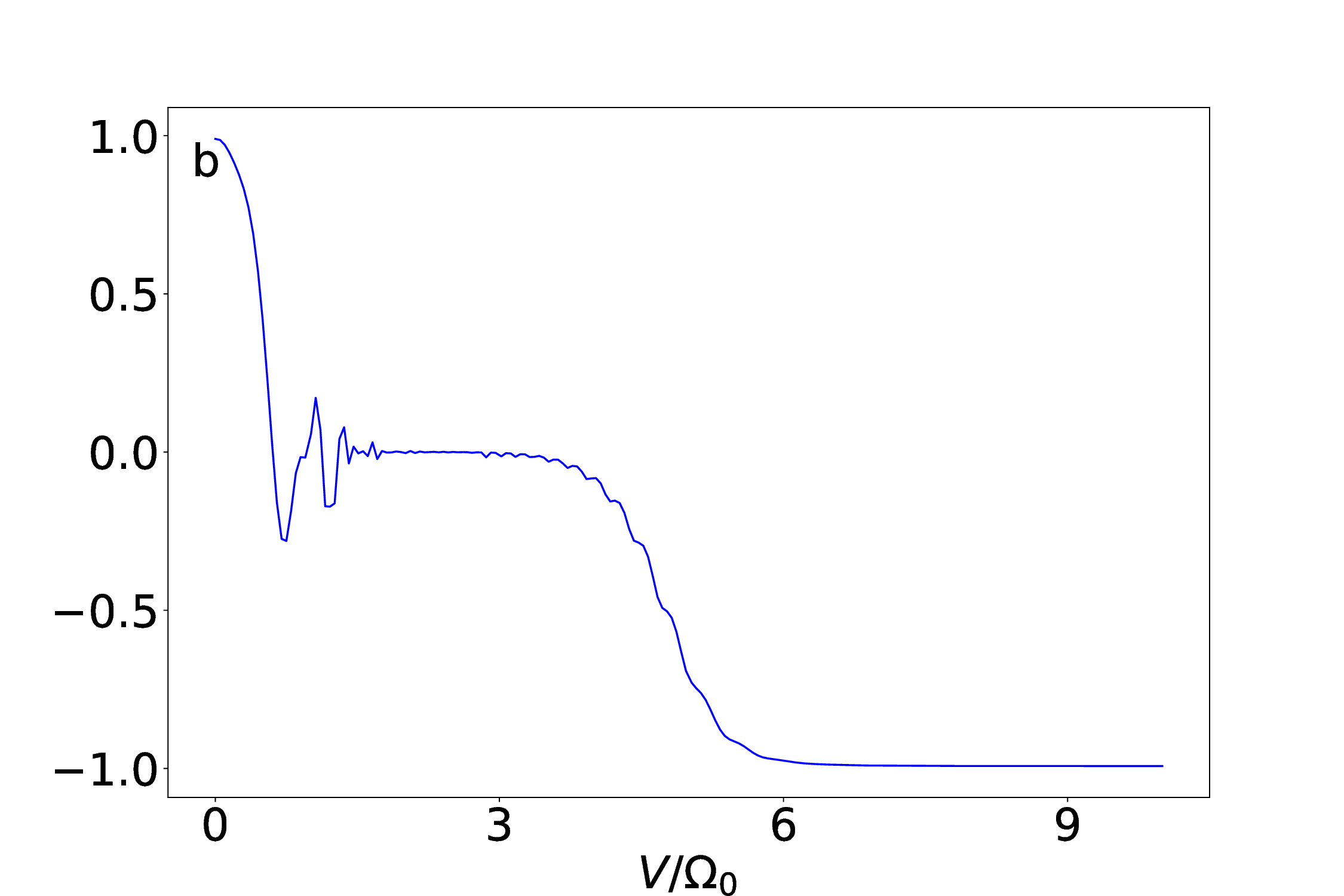}
     \caption{Real part of the amplitude of the target atom state $\ket{A}$ on the completion of the d-STIRAP process is plotted for two different initial states of the control atom. (a) When the control atom is at $\ket{0}$, for $\Omega_c/\Omega_0>2$, population transfer is blocked $>99\%$, i.e. real part of the amplitude does not change. (b) When initially control atom is in $\ket{1}$, real part of the amplitude of $\ket{A}$ changes as increasing $V$, while we fix $\Omega_c=3$.   }
     \label{fig4}
 \end{figure}

First, we analyze the evolution of amplitudes of the computational basis states as a function of gate parameters after the completion of the two-qubit gate operation process. In the Fig.\ref{fig4}a, we notice that for $\Omega_C/\Omega_0>2$ we find the highest population transfer blockage for the initial state $\ket{0A}$, i.e., amplitude does not change, and Fig.\ref{fig4}b shows the change of amplitude of $\ket{1A}$ with increasing $V$.

\subsection{Fidelity}\label{fid}
 \begin{figure}
     \centering
     \includegraphics [width=9cm]
     {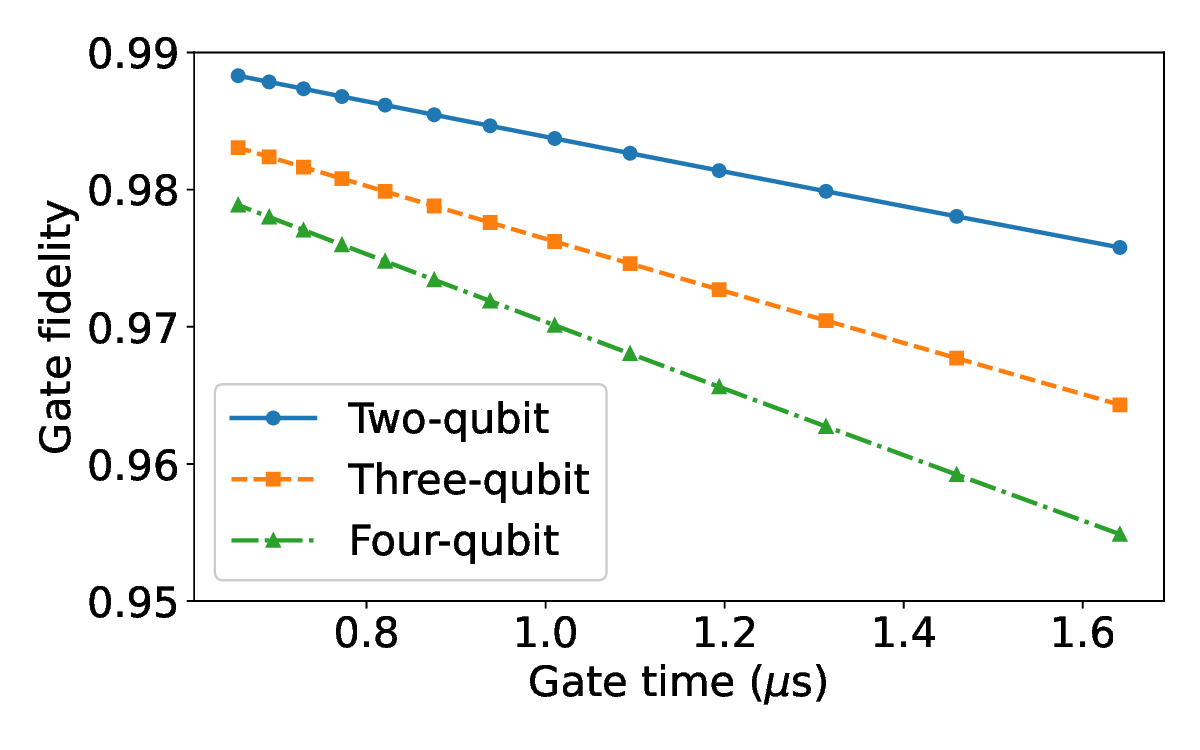}
     \caption{Gate fidelities for two, three, and four qubit cases are plotted varying the gate time.}
     \label{fig5}
 \end{figure}
 For numerical illustration, we consider the $^{133}$Cs atom. For the control atom, we choose two ground state hyperfine sublevels of $6S_{1/2}(F=3,4)$ as the qubit states. For the target atom, we take three Zeeman sublevels $m_F=-1,0,1$ from the seven sublevels of $F=3$ as the ground states, and from these three, two ground states are treated as qubit states. $126S$ state with lifetime 540 $\mu$s \cite{Saffman2020_SymmetricRydbergCZ} is chosen as the Rydberg states for both control and target atoms.  We choose the intermediate states as $7P_{3/2}$ and $7P_{1/2}$ with the corresponding lifetimes 137.54 ns and 165.21 ns, respectively \cite{Toh2019_7pCesiumLifetimes}.

The average gate fidelity can be evaluated using a trace-preserving, operator-based expression \cite{nielsen2002simple}:
\begin{equation}\label{Efid}
   \bar{F}(\hat{\mathcal{O}},\epsilon)
   = \frac{\sum_j \mathrm{Tr}\!\left[\hat{\mathcal{O}}\,\hat{\mathcal{O}}_j^\dagger\,\hat{\mathcal{O}}^\dagger\,\epsilon(\hat{\mathcal{O}}_j)\right] + d^2}{d^2(d+1)} ,
\end{equation}
where $\{\hat{\mathcal{O}}_j\}$ represents the complete set of tensor products of $N$-qubit Pauli operators. 
For instance, in the two-qubit case, the operator set is $
(\hat{I}\otimes\hat{I},\; \hat{I}\otimes\hat{\sigma}_x,\; \ldots,\; \hat{\sigma}_z\otimes\hat{\sigma}_z).
$ Here, $\hat{\mathcal{O}}$ denotes the ideal universal quantum gate, 
$\epsilon$ is the trace-preserving quantum operation realized by the proposed gate implementation, 
and $d = 2^N$ denotes the Hilbert space dimension of an $N$-qubit system. In the presence of the spontaneous emission of Rydberg atoms, the evolution of the whole system can be governed by the master equation
\begin{equation}
    \dot{\hat{\rho}}=-i[\hat H, \hat \rho]+ \sum_{i=0}^1\mathscr{D}_1[\sigma_i]\hat{\rho}+\mathscr{D}_2[\alpha]\hat{\rho}+\sum_{j=A}^C\mathscr{D}_3[\beta_j]\hat{\rho}+\sum_{k=B}^C\mathscr{D}_4[\beta_k']\hat{\rho},
\end{equation}
where $\mathscr {D}[\hat a]\hat \rho=a\hat \rho\hat a^\dagger-(\hat\rho\hat a^\dagger \hat a+\hat{a}^\dagger\hat a\hat \rho)/2$. The decay channels are defined as  $ \sigma_i=\sqrt{\gamma_r/2}\ket{i}\bra{r}$, $\alpha=\sqrt{\gamma_R}\ket{e_1}\bra{R} $, $\beta_j=\sqrt{\gamma_e/2}\ket{j}\bra{e_1}$ and  $\beta_k'=\sqrt{\gamma_e/2}\ket{k}\bra{e_2}$. We analyze the average gate fidelity by varying the gate operation time in fig.\ref{fig5}.
 
 \subsection{Error analysis}
\label{errors}
 \subsubsection{Rabi error}
\begin{figure}
     \centering
     \includegraphics [width=8.5cm]
     {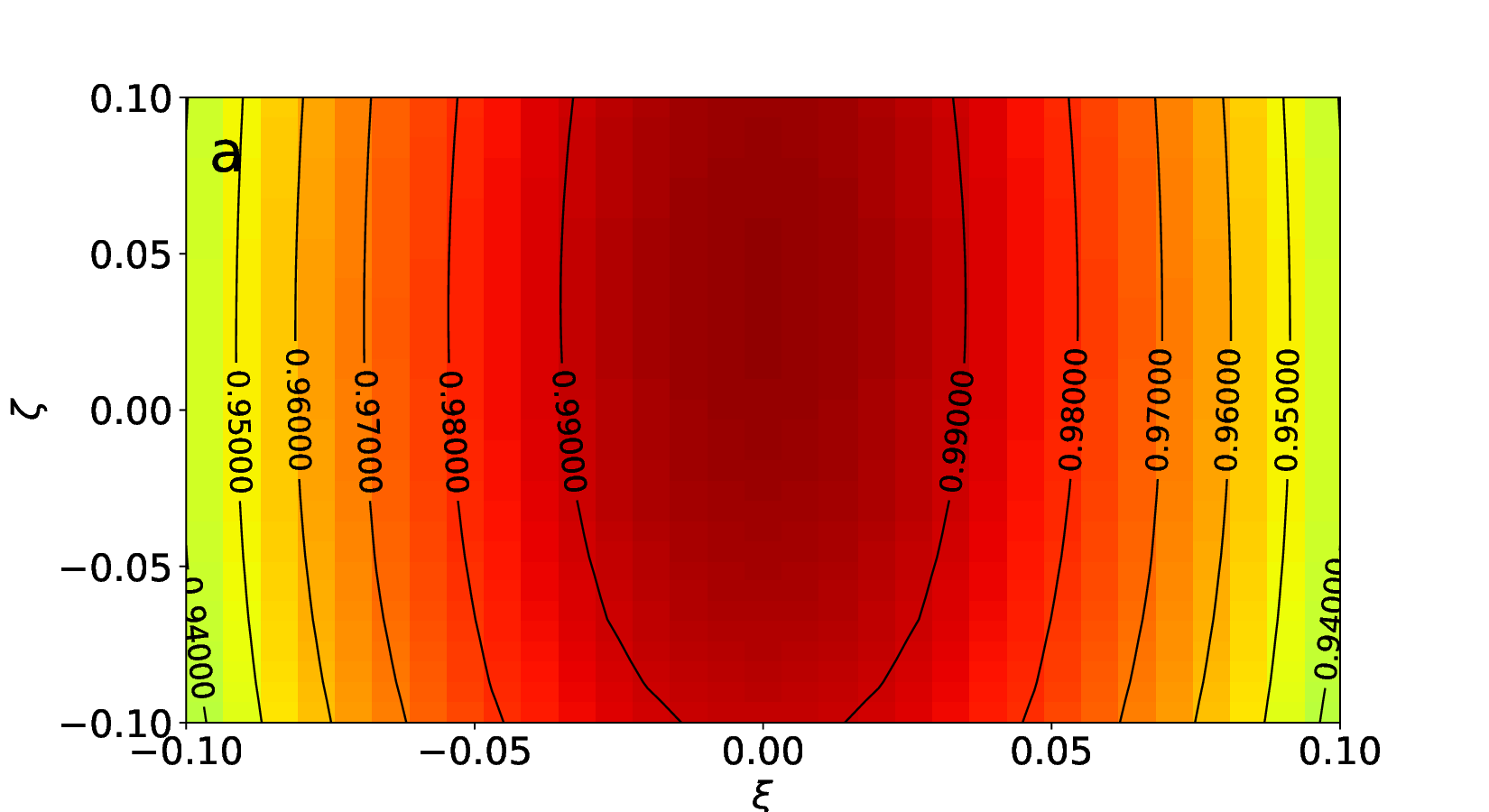}
     \includegraphics [width=8.5cm]{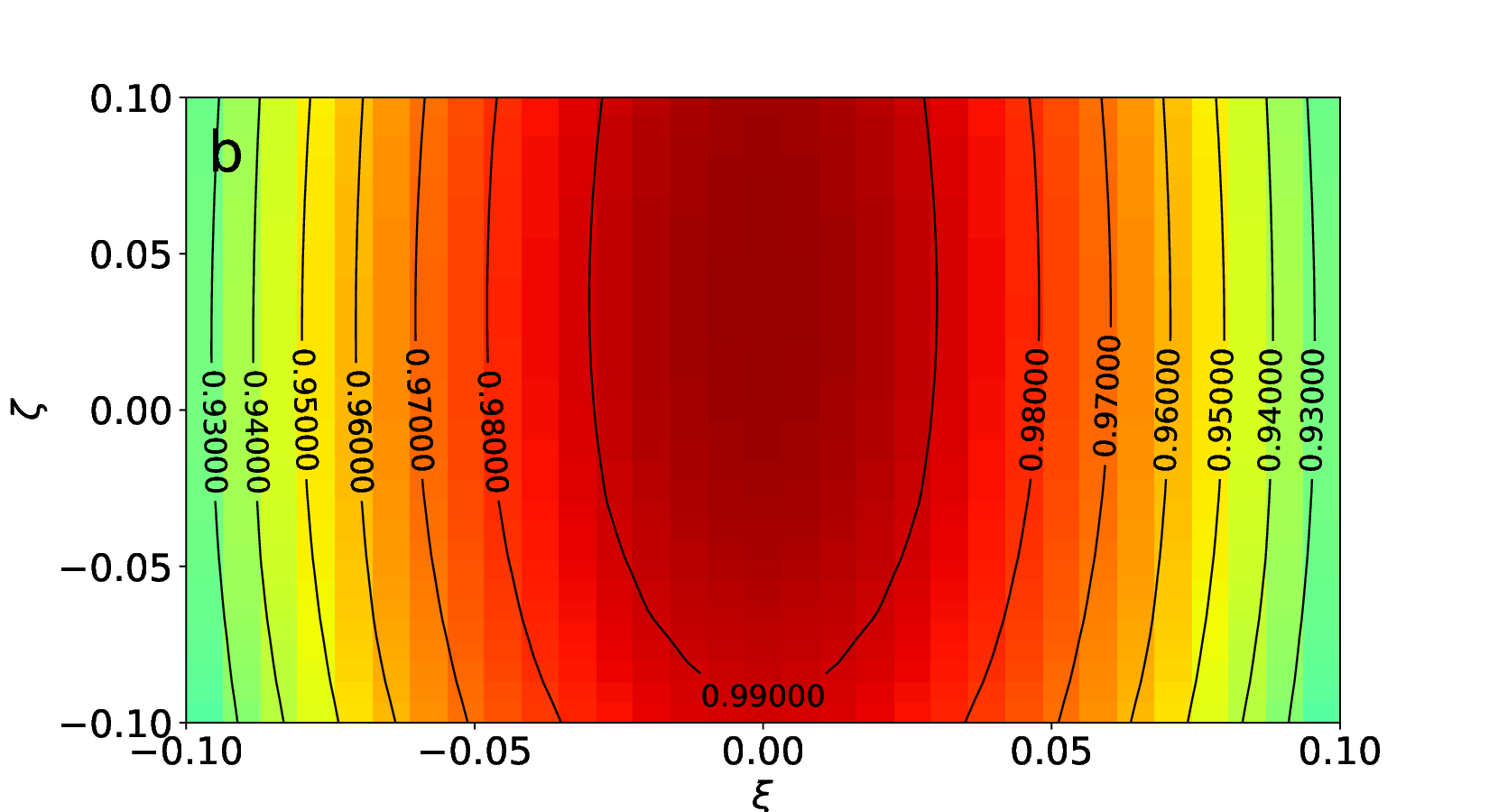}\\
     \includegraphics [width=8.5cm]
     {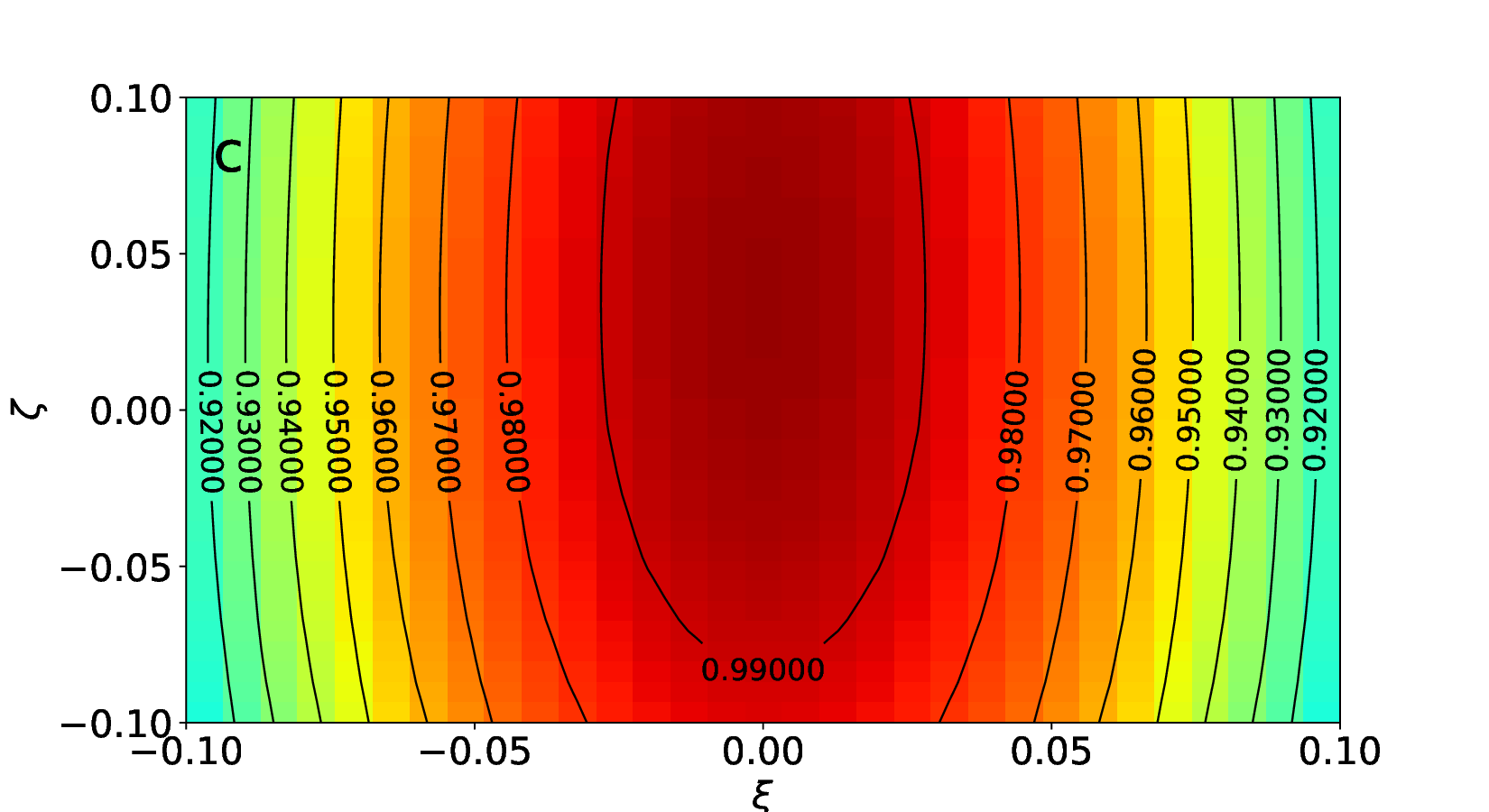}
     \includegraphics [width=8.5cm]{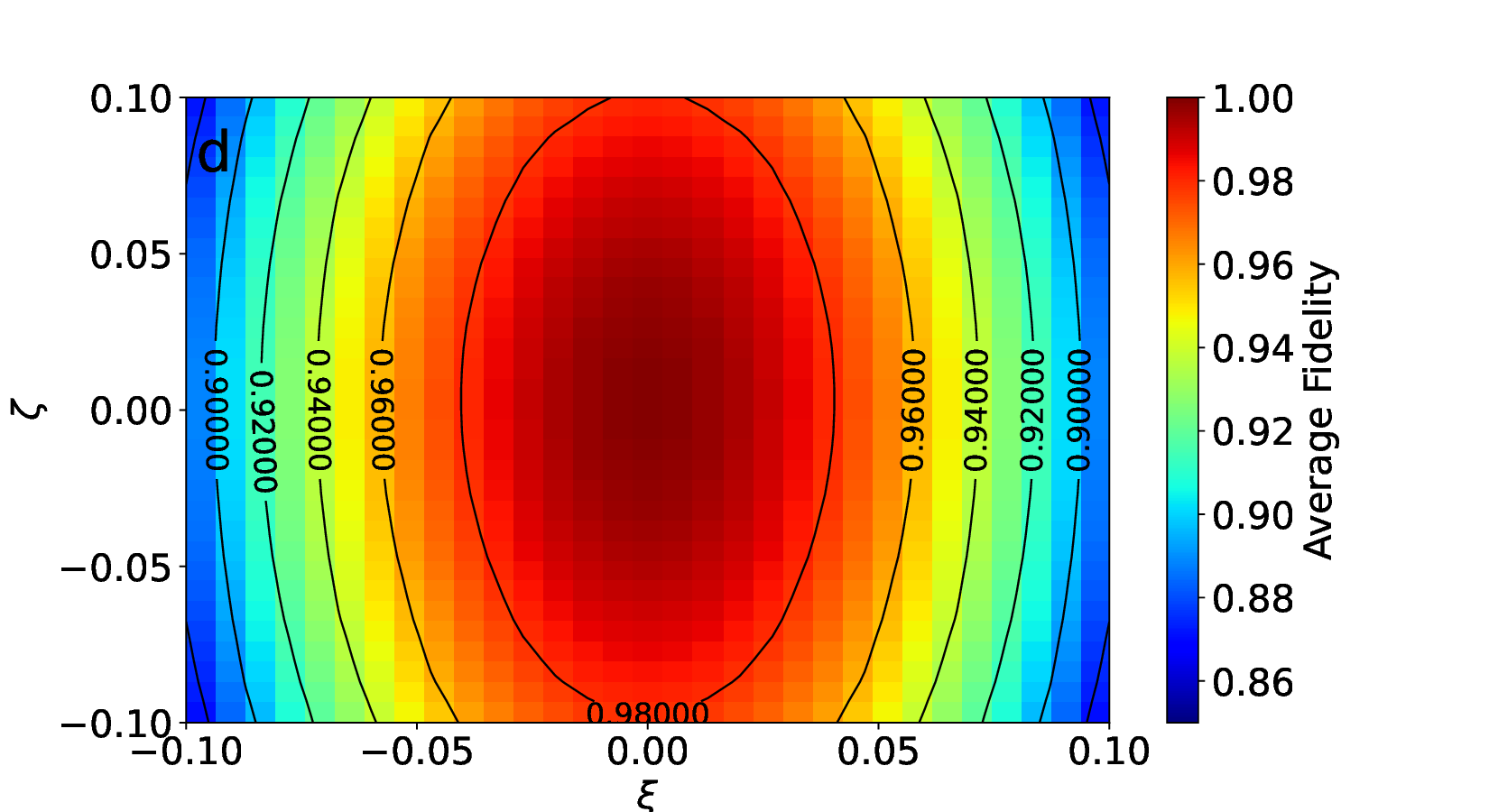}
     \caption{(a), (b), (c) show robustness of our proposed two-, three-, and four-qubit controlled phase gates against the Rabi error, respectively. (d) shows the change in gate fidelity of the conventional EIT-based CNOT gate for the Rabi error. }
     \label{fig6}
 \end{figure}

 When designing quantum logic gates theoretically, it is customary to assume that the Rabi frequency remains constant throughout the gate operation. In realistic experimental settings, however, fluctuations in the laser intensity inevitably introduce deviations in the Rabi frequency, which in turn lead to systematic errors. To capture this effect and to analyze the robustness of our gate protocol, we model the Hamiltonian in the presence of such imperfections. Specifically, in the first and third steps, the control Hamiltonian is modified as
$ H^{\mathrm{error}}_c(t) = (1+\xi) H_c(t),$ 
and in the second step, the target Hamiltonian becomes
$H^{\mathrm{error}}_t(t) = (1+\zeta) H_t(t),$
where $\xi$ and $\zeta$ are error parameters that quantify the relative deviations in the Rabi frequency due to laser intensity fluctuations. This formulation allows us to explicitly investigate how such systematic errors influence the gate performance and to demonstrate the intrinsic robustness of the proposed gate scheme against variations in the driving field strength. In the Fig.\ref{fig6}a, b, c, we plot the average gate fidelity of our proposed two, three, and four-qubit controlled phase gate, respectively, varying the Rabi frequency of the control and target atom up to $10\%$ using the formula described in Eq.\ref{Efid}. We notice that, especially for the Rabi error in the target atom, our gates are superbly robust. We also compare that our proposed scheme shows more resilience than the conventional EIT-based CNOT gate \cite{Mueller2009Mesoscopic} (see fig.\ref{fig6}d).

 \subsubsection{Blockade error}
\begin{figure}
     \centering
     \includegraphics [width=5.9cm]
     {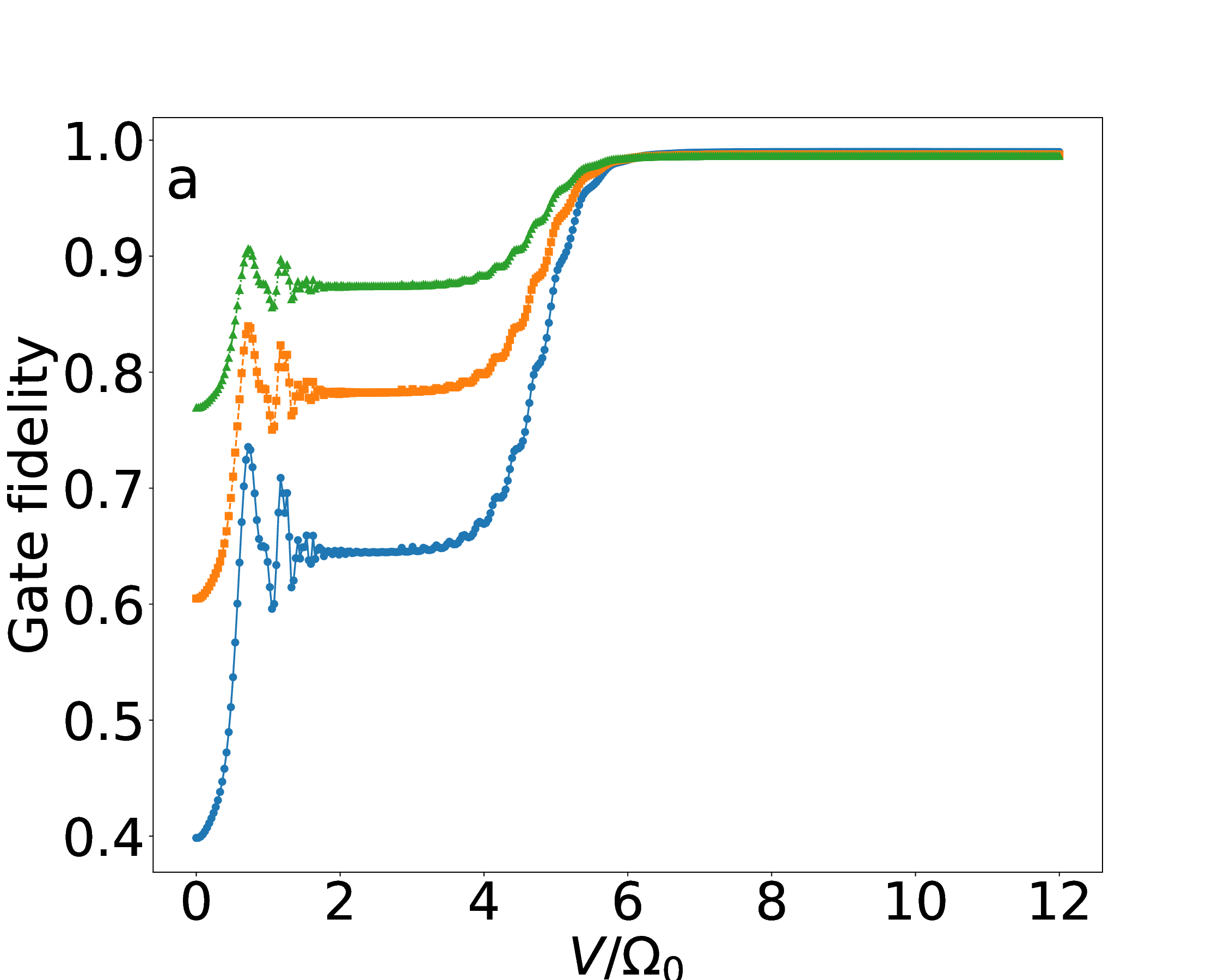}
     \includegraphics [width=5.9cm]{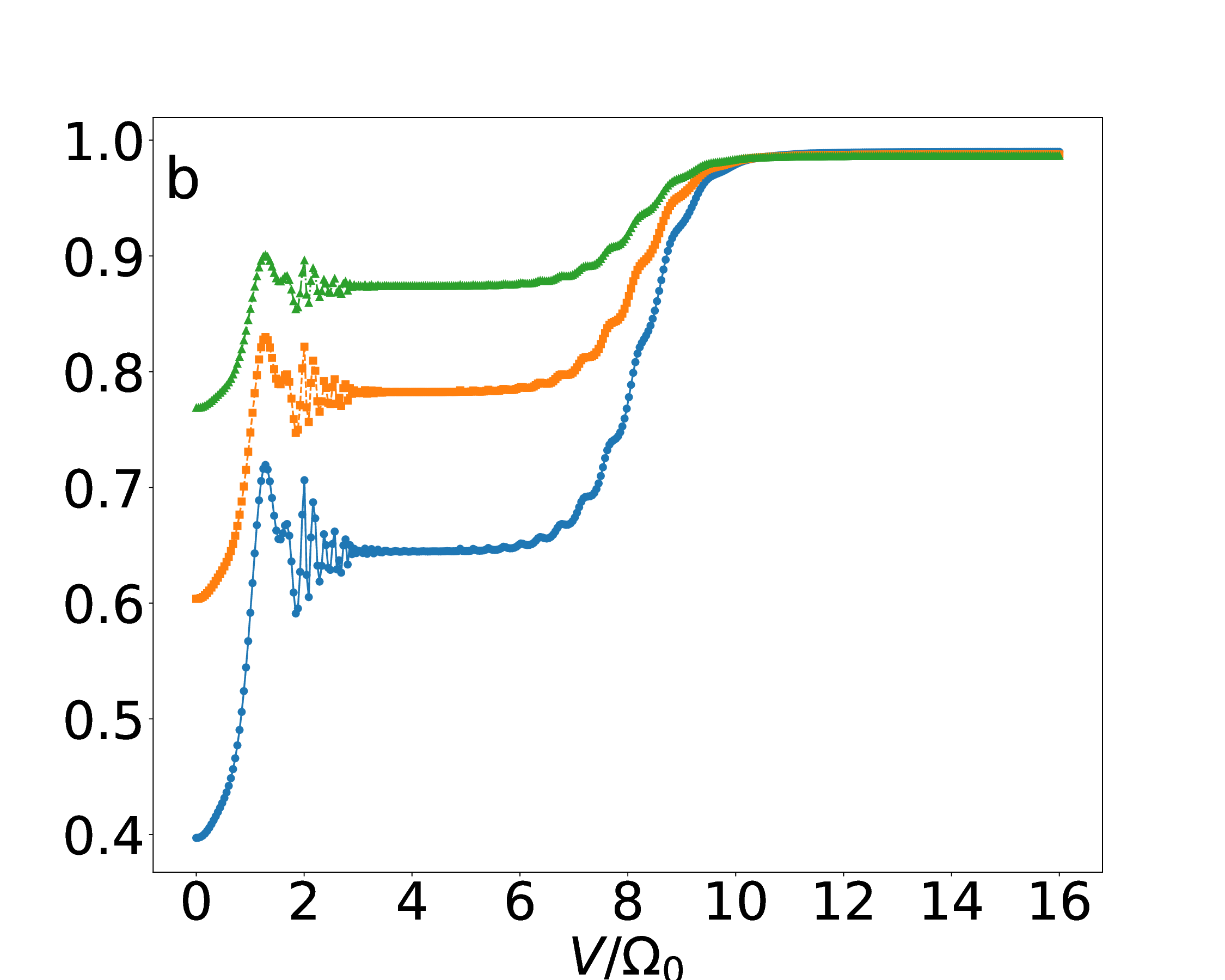}
     \includegraphics [width=5.9cm]
     {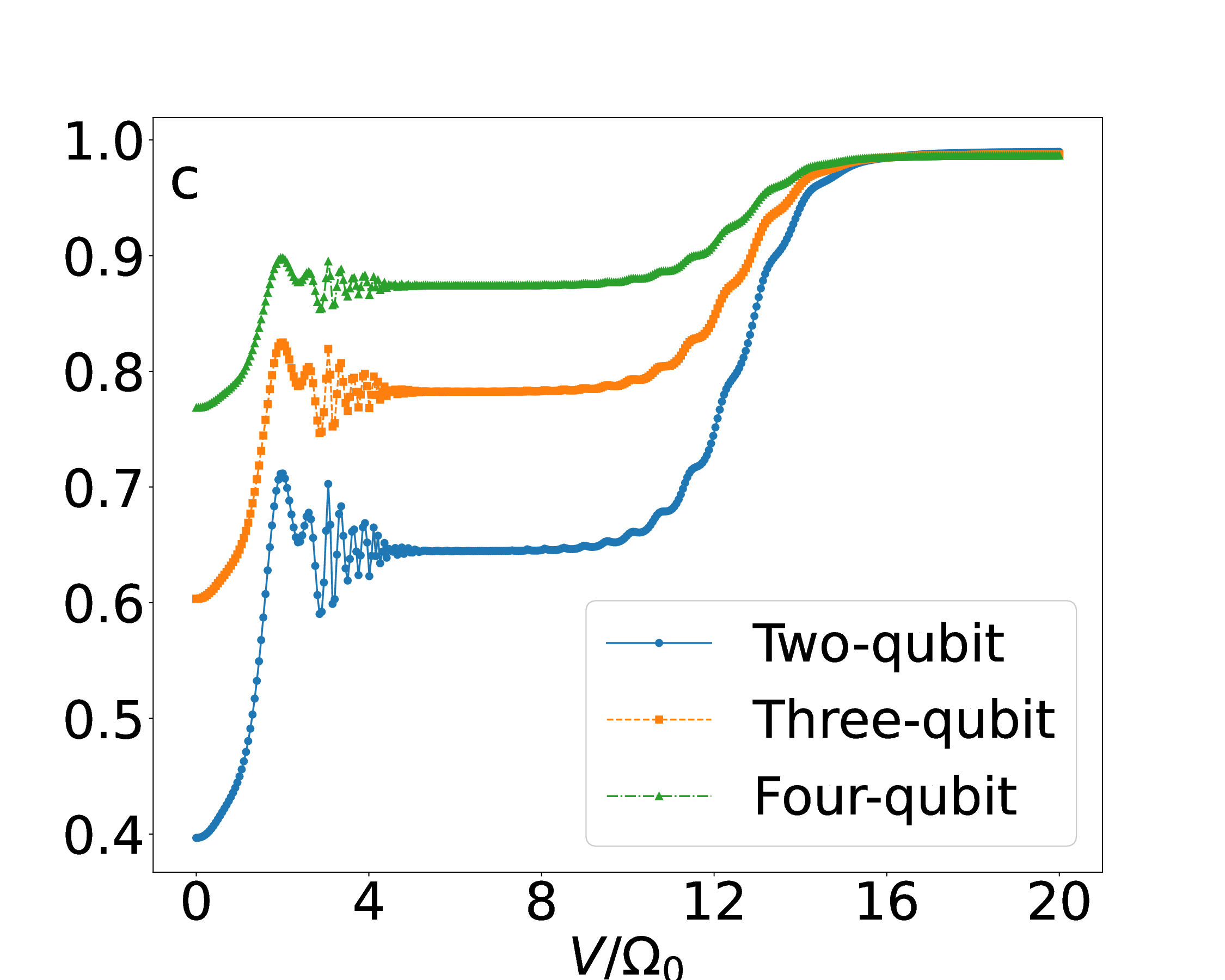} 
     \caption{(a), (b), (c) show that maximum gate fidelity can be achieved for $V = 6(10,15)\Omega_0$, i.e., $V = 2(2.5,3)\Omega_c$ while taking  $\Omega_c = 3(4,5)\Omega_0$ respectively.}
     \label{fig7}
 \end{figure}
In Rydberg atom quantum gates, blockade error arises when the Rydberg blockade mechanism is not perfect. Ideally, excitation of one atom to a Rydberg state should shift the energy levels of nearby atoms strongly enough to completely prevent their excitation. However, in practice, the interaction strength is finite, and the laser Rabi frequency may be comparable to or larger than the blockade shift. As a result, there is a small probability that two atoms are simultaneously excited to Rydberg states. The blockade error typically scales as $(\Omega_c/V)^2$. For conventional Rydberg blockade-based gates, the interaction strength must significantly exceed the Rabi frequency in order to suppress blockade errors. In our scheme, we demonstrate that maximum gate fidelity can be achieved when the interaction strength satisfies  $V = 6(10,15)\Omega_0$, i.e., $V = 2(2.5,3)\Omega_c$ with $\Omega_c = 3(4,5)\Omega_0$, as illustrated in Fig.~\ref{fig7}. This result highlights the robustness of our gate against the blockade error.

 \subsubsection{Positional fluctuation error}
The vdW interaction varies with the interatomic distance following the equation $V=(C_6/l^6)$. The vdW coefficient ($C_6$) can be calculated for Cs using  the following formula \cite{Singer2005}
\begin{equation}
    C_6=\mathcal{N}^{11}(10.64-0.6294\mathcal{N}+2.33\times10^{-3}\mathcal{N}^2) \ \text{a.u}.
\end{equation} 
where $\mathcal{N}$ is the principal qauntum number. For instance we take $\mathcal{N}=126$ and $\Omega_0/2\pi=44$MHz. For $\Omega_c=3\Omega_0$ and $l=6\ \mu$m we get $V=94\Omega_c$ . From Fig.\ref{fig7}a it is clear that for $V\gg2\Omega_c$ some fluctuation in the position of the atoms during the experiment will not affect the fidelity of our proposed gates.

 \subsection{Grover's search algorithm}
\label{grover}
\begin{figure}
     \centering
     \includegraphics [width=12cm]
     {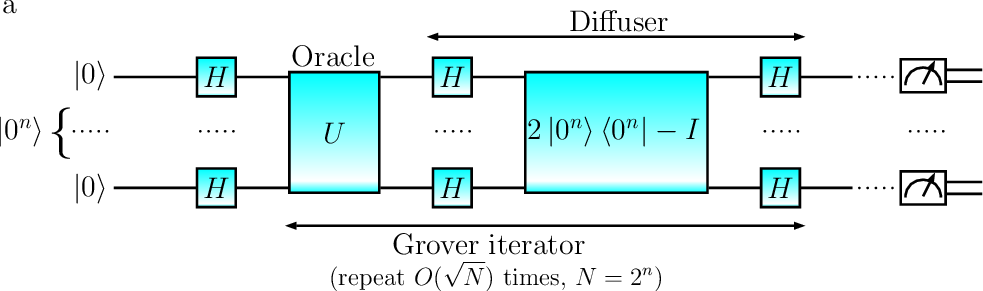}\\
     \includegraphics [width=8cm]
     {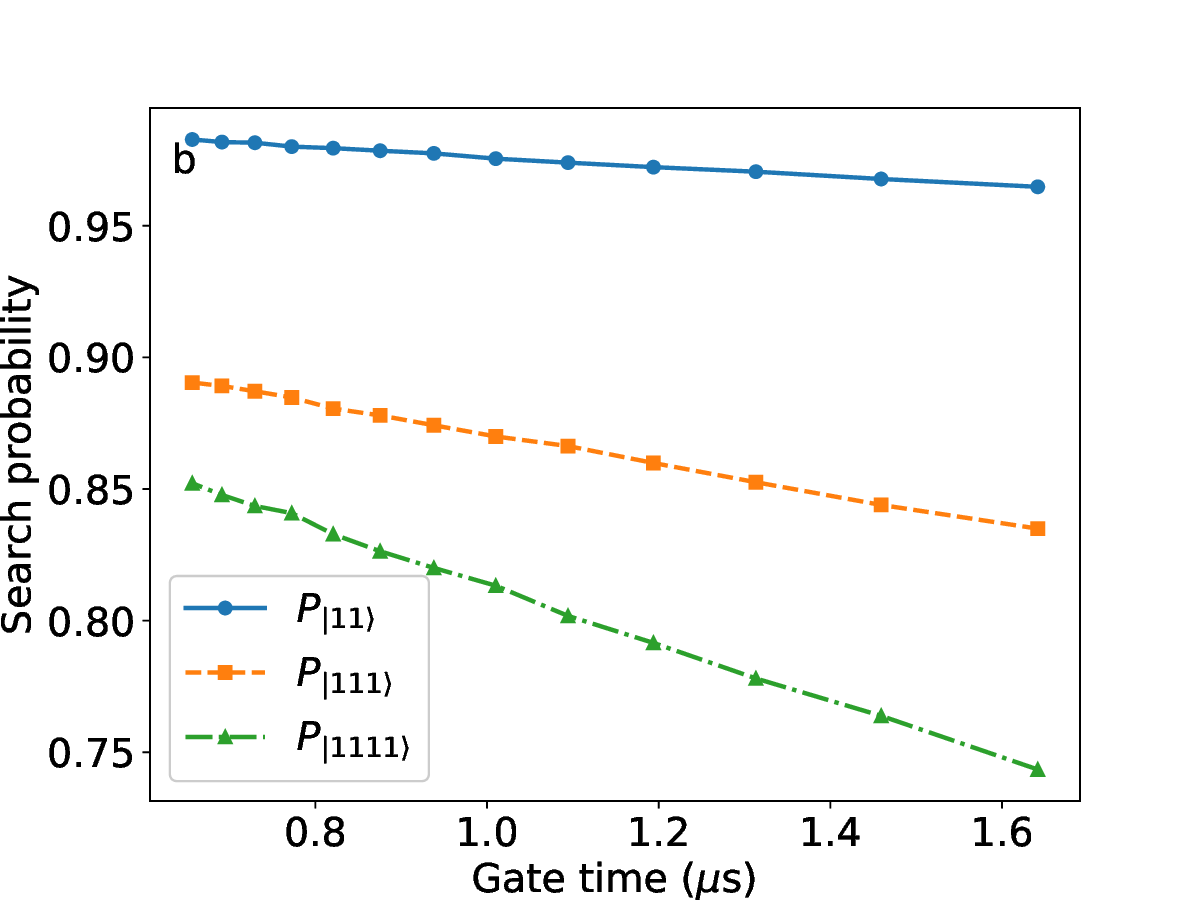}
     \caption{(a) The circuit diagram for simulating n-qubit Grover's search algorithm. (b) shows the change in search probability of $\ket{11}$, $\ket{111}$ and $\ket{1111}$ for varying the gate time of two-, three- and four-qubit controlled phase gates.}
     \label{fig8}
 \end{figure}
To demonstrate the computational power and practical relevance of our proposed multi-qubit gates, we implement Grover’s quantum search algorithm using our proposed quantum gates. Grover’s algorithm achieves a quadratic speedup compared to its classical counterpart when solving unstructured search problems~\cite{Grover1996, Long2002, Liu2022G}. In a classical scenario, identifying a marked item within a database of size $N$ requires, on average, $\mathcal{O}(N)$ oracle queries. By contrast, Grover’s algorithm leverages quantum superposition and amplitude amplification to locate the marked element with only $\mathcal{O}(\sqrt{N})$ oracle queries, making it one of the most prominent demonstrations of quantum speedup.  As a concrete illustration, we simulate two-, three-, and four-qubit versions of Grover’s algorithm, targeting the marked states $\ket{11}$, $\ket{111}$, and $\ket{1111}$, respectively. These correspond to searches within Hilbert spaces of dimensions four, eight, and sixteen. Our simulations explicitly employ the proposed multi-qubit gates to construct both the oracle and diffusion operators, thereby highlighting their effectiveness in nontrivial quantum algorithms. In Fig.\ref{fig8}b, we present the evolution of the success probability of finding the marked state as a function of the individual gate operation time. The observed amplification dynamics confirm that our gate scheme faithfully reproduces the expected Grover search behaviour, while also underscoring its scalability to larger Hilbert spaces.

 \section{conclusions} In conclusion, we have proposed and demonstrated two-qubit geometric phase gates based on dark states and a pair of d-STIRAP pulse sequences. We have generalized the protocol to implement n-qubit controlled phase gates. For example, we have constructed three- and four-qubit gates in one- and two-dimensional atomic configurations, respectively. The system is designed in such a way that no extra laser is required to be applied on the
target atom while scaling up the system from two- to multi-qubit quantum gates. Using realistic parameters, we have demonstrated that these gates can achieve high fidelities ($98\%$ to $99\%$) when the atoms are positioned at sufficiently large separations ($l=6\ \mu$m) to allow individual addressing. Furthermore, we have carried out a detailed analysis of the gate performance under various systematic errors, including Rabi frequency fluctuations, blockade imperfections, and positional uncertainties, and found that the proposed multi-qubit gates exhibit strong resilience against all of these errors. Finally, to highlight the computational utility of our scheme, we have simulated Grover’s search algorithm with two, three, and four qubits using the proposed gates and obtained high success probabilities. These results establish our approach as a scalable and experimentally feasible route for realizing multi-qubit gates.
\bibliographystyle{apsrev4-2.bst}
\bibliography{ref}
\end{document}